\documentclass[apj]{emulateapj}
\usepackage{amsmath}
\usepackage{graphicx}
\usepackage{natbib}
\usepackage{epsfig}

\usepackage{url}
\usepackage{indentfirst}
\usepackage{color}

\begin{document}
\nocite{*}
\title{A multi-wavelength study of PSR J1119$-$6127 after 2016 outburst}
\author{H.-H. Wang\altaffilmark{1}, L.C.-C. Lin\altaffilmark{2}, S. Dai\altaffilmark{3}, J. Takata\altaffilmark{1}, K.L. Li.\altaffilmark{2}, C.-P. Hu\altaffilmark{4, \textdagger}, 
X. Hou \altaffilmark{5,6}}
\email{wanghh@hust.edu.cn, lupin@unist.ac.kr, takata@hust.edu.cn}
\altaffiltext{1}{Department of Astronomy, School of physics, Huazhong University of Science and Technology, Wuhan 430074, China}
\altaffiltext{2}{Department of Physics, UNIST, Ulsan 44919, Korea}
\altaffiltext{3}{CSIRO Astronomy and Space Science, Australia Telescope National Facility, Box 76 Epping NSW 1710, Australia}
\altaffiltext{4}{Department of Astronomy, Kyoto University, Oiwakecho, Sakyoku, Kyoto 606-8502, Japan}
\altaffiltext{5}{Yunnan Observatories, Chinese Academy of Sciences, Kunming, 650216, China}
\altaffiltext{6}{Key laboratory for the Structure and Evolution of Celestial Objects, Chinese Academy of Sciences, Kunming, 650216, China}
\altaffiltext{\textdagger}{JSPS International Research Fellow}

\begin{abstract}
   PSR~J1119$-$6127, a high-magnetic field pulsar detected from radio to high-energy wavelengths, underwent a magnetar-like outburst beginning on July 27, 2016. In this paper, we study the post-outburst multi-wavelength properties of this pulsar from the radio to GeV bands and discuss its similarity  with the outburst of the magnetar XTE~J1810$-$197.   In phase-resolved spectral analysis of 0.5--10\,keV X-ray data collected in August 2016, the on-pulse and off-pulse spectra are both characterized by two blackbody components and also require a power-law component similar to the hard X-ray spectra of magnetars.  This power-law component is no longer distinguishable in data from December, 2016. 
  We likewise find that there was no substantial shift between the radio and X-ray pulse peaks after the 2016 X-ray outburst.  
  The gamma-ray pulsation after the X-ray outburst is confirmed with data taken after 2016 December and the pulse structure and phase difference between the gamma-ray and radio peaks ($\sim$0.4 cycle) are also consistent with those before the X-ray outburst. 
 These multi-wavelength observations suggest that the re-configuration of the global magnetosphere after 2016 magnetar-like outburst at most continued for about six months.   
  We discuss the evolution of the X-ray emission after the 2016 outburst  with  the untwisting  magnetosphere model.
 \end{abstract}
\section{introduction}
An isolated pulsar is a rapidly rotating and highly-magnetized neutron star with a spin period of  \textbf{$\sim$ 1\,ms--$\sim$ 10\,s }and a surface
dipole field of $B_s \sim 10^{8-15}$G. Magnetars form a subclass of pulsars with bright and variable emission in the X-ray and gamma-ray band \citep{Kaspi17}.  Since the observed radiation from the magnetars  exceeds the  spin-down power of the pulsar, it has been
  suggested that decay of an ultra-high magnetic field that exceeds the  critical
  magnetic field of {$B_{c}$=$m_{e}^2c^3/e\hbar\sim4.4\times 10^{13}$}\,G provides the energy source of the emission \citep{Thompson96}.
  The most remarkable feature of magnetars is their short, bright bursts, likely powered by a sudden release of energy from the star's magnetic field \citep{WT2006,Ng2011,Pons2012}.  Some bursts are accompanied by longer duration outburst states.  It has been confirmed that some rotation powered pulsars (RPPs) with a high magnetic field strength
($B_s>10^{13}$\,G, hereafter high-$B$ pulsars) also show magnetar-like X-ray outbursts \citep{Livingstone2010,Archibald2016}, and these high-$B$ pulsars could provide a connection between magnetars and the usual RPPs.

Observed X-ray outbursts of magnetars are frequently accompanied by a glitch.
Glitches are a sudden change in the spin frequency ($f$) and spin-down rate {($\dot{f}$)}, and the radio observations have revealed  $\sim $180 glitching pulsars  (Espinoza et al. 2011). The distribution
is bimodal in glitch size ($\triangle f/f$)  and is divided into large (small) glitches with $\triangle f/f \ge 10^{-7}$ ($\le 10^{-7}$).  Magnetar glitches are large \citep{DKG2008}, and the main difference with respect to glitches of normal pulsars is the presence of an accompanying X-ray outburst and pulse shape change \citep{Kaspi03,Woods04}, typically absent for glitches of normal pulsars.

PSR~J1119$-$6127 is a high-$B$ pulsar with an inferred polar strength of $B_s \sim 6.4\times 10^{19}\sqrt{P\dot{P}}$ =$8\times 10^{13}$\,G \citep{Camilo2000}, with a spin period $P=0.407$\,s and a period time derivative $\dot{P}=4\times10^{-12}{\rm s\,s^{-1}}$, giving a characteristic age and spin-down power of the pulsar $\tau_{c}$=$P/2 \dot{P}$=$1.6$\,kyr and  \textbf{$\dot E_{sd}$=$2.3 \times 10^{36}$\,${\rm erg~s^{-1}}$,} respectively.
This pulsar was discovered in the Parkes multibeam pulsar
survey \citep{Camilo2000} and it is likely associated with the supernova remnant G292.2-0.5 \citep{Crawford01}
at a distance of 8.4 kpc \citep{CMC2004}. Pulsed emission was also detected in the  X-ray and gamma-ray
bands \citep{GS03,Parent2011}. PSR~J1119$-$6127 glitched in 1999, 2004, and 2007 \citep{Camilo2000,WJE2011,Anton2015}. The glitches of 2004 and 2007 exhibited a recovery of
the spin-down rate toward the pre-glitch level, but that of 2007 showed an over-recovery that  continued to evolve
on a time scale of years \citep{Anton2015}. Unusually, the 2007 glitch was also accompanied by an X-ray outburst, with a temporary change in the radio pulse profile from single- to double-peaked \citep{WJE2011}.

The  \emph{Fermi} Gamma-ray Burst Monitor (GBM) and \emph{Swift} Burst Alert Telescope (BAT) both triggered on magnetar-like X-ray outburst{s} of PSR~J1119$-$6127 on 2016 July 27 UT 13:02:08 \citep{Younes2016} and on 2016 July 28 UT 01:27:51 \citep{Kennea2016}. 
\citet{Gogus2016} identified 13 short X-ray bursts between
2016 July 26 and 28, with an estimate of the total energy released of $E\sim 10^{42}\,{\rm erg}$.  The pulsar also underwent
a large glitch immediately after the start of the 2016 outburst \citep{Archibald2016}. Following the X-ray outburst, monitoring of this source was carried out in radio and X-ray bands.

\begin{figure*}
  \includegraphics[scale=0.4]{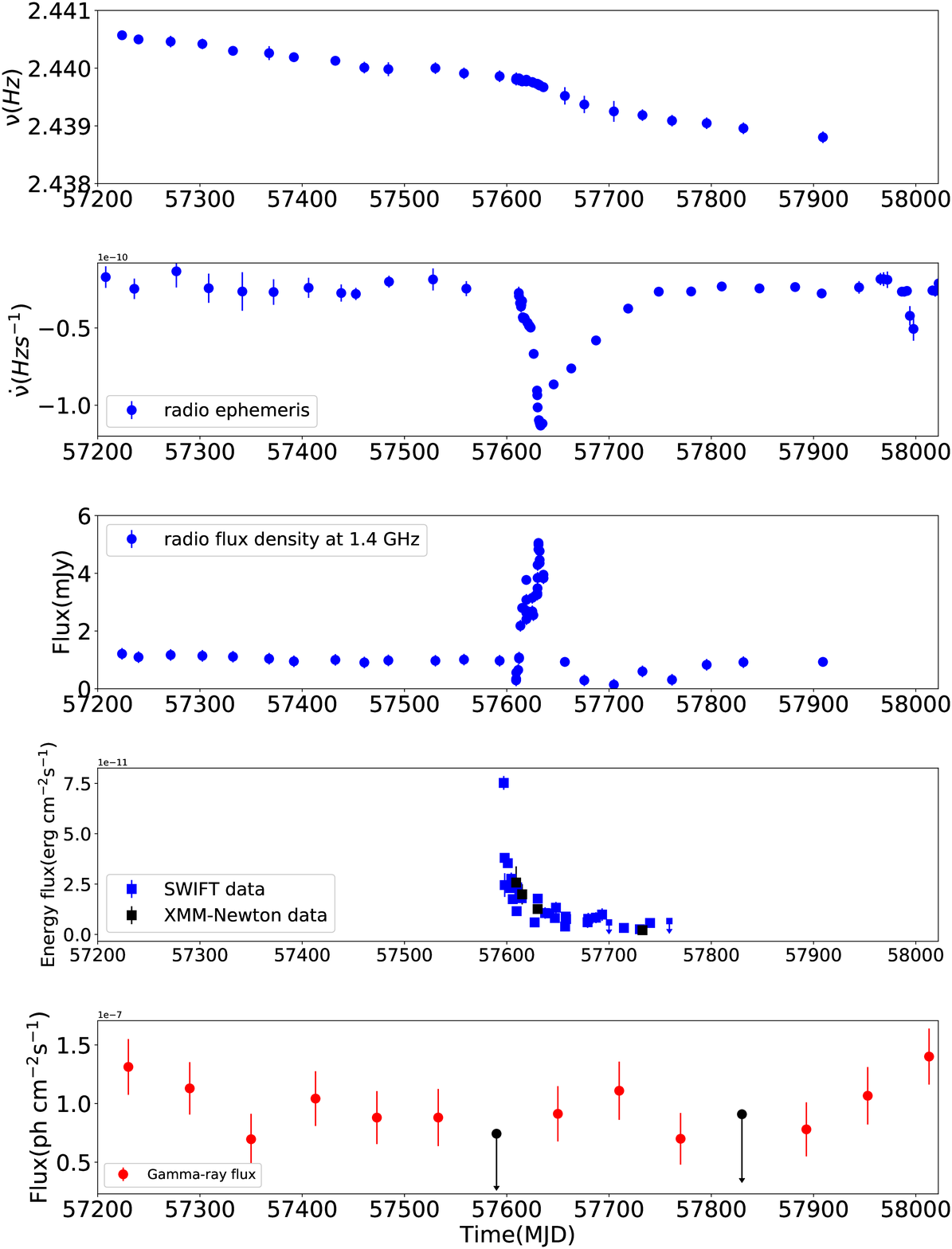}
  \caption{\label{fig:allband}From top to bottom, the panels show evolution of the spin frequency, spin-down rate,radio flux density, X-ray flux and  $ >$0.1\,GeV flux of PSR~J1119$-$6127, respectively. In the bottom panel, each point represents the flux $ >$0.1\,GeV assessed from a 60 day accumulation of \emph{Fermi} data, and the black arrow indicates a 3$\sigma$ flux upper limit.}
 \label{glc}
\end{figure*}

Following the glitch, the pulsed radio emission disappeared for about two weeks,
and the reappearance of the pulse profile exhibited
a multi-component structure at 2.3\,GHz and a single peak at 8.4\,GHz \citep{Majid2017, Dai2018}.
After the glitch, the spin-down rate rapidly increased by a factor of 5--10 on 2016 September 1 (MJD $\sim$57632)  before recovering toward the pre-burst rate over the following three months \citep{Dai2018,Archibald2018,Lin2018}.  Moreover, the radio flux increased by a factor of ten, and then started decreasing  at around MJD 57632, which showed an obvious correlation with the evolution of the spin-down later after the outburst (Figure \ref{fig:allband}).  
During the rapid evolution of the spin-down rate, the radio pulse profile changed twice, on 2016 August 12 (MJD $\sim$57612) and 29 (MJD $\sim$57630).
We note that 
the increase of the spin-down rate became faster at around August 12, and the spin-down rate began recovery to pre-glitch levels around August 29. 
\citep{Archibald2017} also identified three short X-ray outbursts coincident with the suppression of the radio flux on 2016 August 30. Many of these trends are depicted in Figure \ref{fig:allband}.
The Radio timing solutions shown in Figure \ref{fig:allband} were also presented in Dai et al. (2018), formed using observations with the Parkes telescope using an observing bandwidth of 256 MHz centered at 1369 MHz.

In X-rays, \citet{Archibald2018} find the luminosity  after the
outburst reaches $L_X\sim 4\times 10^{35}{\rm erg~s^{-1}}$,
which corresponds to $\sim$17\% of the spin-down power,
suggesting the energy source of the emission is the dissipation of
magnetic energy.  Moreover, they also report a hardening of the spectrum and an increase in
the 0.7--2.5\,keV (2.5--10\,keV) pulsed fraction from $\sim 38$\% ($<10$\%) before the outburst to $\sim$71\% ($\sim$56\%).  \citet{Lin2018} investigated the high-energy emission features after the X-ray
outburst, and they report that the X-ray emission in 2016 August can be described by  two thermal components of different spatial sizes plus
a power-law component with a photon index $\Gamma\le 1$ (see also \citet{Archibald2018}). 
This hard photon index above 10\,keV is similar to magnetar emission \citep{Enoto2017}, rather than that of normal pulsars,
for which a typical photon index $\sim$1.5 is thought to originate from synchrotron radiation from
secondary electron/positron pairs created by the pair-creation process.
It is likely that the X-ray outburst of PSR~J1119$-$6127 and emission after its outburst operated under a magnetar-like process. \citet{BSM2017} report that a single power-law model with a
photon index of $\sim$2 is sufficient to model the X-ray spectrum taken from the \emph{Chandra} observation at the end of 2016 October,
while \citet{Lin2018} and \citet{Archibald2018} both fit the \emph{XMM}-Newton data in 2016 December with a composite model of a single power-law plus two
blackbody and one-temperature blackbody component, respectively. These multi-wavelength observations help constrain the theoretical
interpretation for the mechanism of the X-ray outburst of the magnetars and high-B pulsars. 

In this paper, we revisit the emission characteristics of PSR~J1119$-$6127 after the 2016 outburst and discuss their implications, based on multi-wavelength (radio/X-ray/GeV gamma-ray) observations. We perform spectral and timing analyses beyond those of previous studies.
 Specifically, we analyze phase-resolved spectra above 10\,keV with \emph{NuSTAR} data and
 we discuss the contribution of hard non-thermal emission to the on-pulse and off-pulse emission. We  examine the relation of the radio and X-ray pulsed peaks, which were aligned with each other observations before the 2016 outburst \citep{Parent2011, Ng2012}. 
We also perform a spectral and  timing analysis of \emph{Fermi} Large Area Telescope (LAT)\footnote{https://fermi.gsfc.nasa.gov/ssc/} data to investigate any possible change of the GeV emission before and after the 2016 outburst.

This paper is organized as follows. In section~\ref{data}, we introduce the reduction of the radio, X-ray and \emph{Fermi} data for this study.
We study the phase-resolved X-ray spectroscopy in section~\ref{xphase} and compare the phase shift between the radio and X-ray pulse peaks. We confirmed the gamma-ray pulsation after the 2016 X-ray outburst and compared the pulse shape and spectrum before and after the outburst in section~\ref{timing}. In section~\ref{discuss},
we discuss the evolution of PSR ~J1119$-$6127 after the outburst within the framework of the twisting magnetosphere model \citep{Beloborodov2009}.

  \section{Data reduction}
  \label{data}

\begin{deluxetable}{cccc}
  \tablecaption{X-ray observations of PSR ~J1119$-$6127
    \label{epheme}}
  \tablewidth{0pt}
  \tablehead{ ObsID& date & instrument &Duration(ks)}
  \multicolumn{3}{c}{\emph{Swift}}\\
  00034632007 &2016 Aug. 09 & XRT/WT &57.6\\
   \multicolumn{3}{c}{\emph{XMM}-Newton} \\
    0741732601 &2016 Aug. 06 &PN &20.1 \\
   0741732701 &2016 Aug. 15 &PN &27.9\\
   0741732801 &2016 Aug. 30 & PN &32.5\\
   0762032801 &2016 Dec. 13 & PN &47.5\\
    \multicolumn{3}{c}{\emph{NuSTAR}} \\
    80102048004&2016 Aug. 05&FPM A/B&127\\
    80102048006&2016 Aug. 14&FPM A/B&170.8\\
    80102048008&2016 Aug. 30&FPM A/B&166.5\\
    80102048010&2016 Dec. 12&FPM A/B&183.3\\
\end{deluxetable}

\subsection{X-ray data} 
\label{xray_analysis}
For X-ray data analysis, we use archival data from the \emph{Neil Gehrels Swift Observatory (\emph{Swift}), X-ray Multi-Mirror Mission (\emph{XMM}-Newton)} and \emph{Nuclear Spectroscopic Telescope Array (\emph{NuSTAR}) }taken in 2016 August and December after the outburst.
Details of observations considered in this study are listed in the Table 1.  For the timing analysis in the X-ray band,  all
events are corrected to the barycenter using the X-ray position (R.A., Decl.) = (11$^{\rm h} 19^{\rm m} 14.26^{\rm s}$,-61$^{\circ}27'49.3''$) and the JPL DE405 solar system ephemeris, and the local timing ephemeris was determined by maximizing the $H$ test statistic value \citep{de2010h}.

For \emph{Swift}, we use data in windowed timing mode with a time resolution of 1.8\,ms carried out by XRT on August 9 and compare the X-ray with radio pulse profiles taken on the same day (section~\ref{peakcom}). 
We extract the source counts from a box region of $15''\times 35''$. Only photon energies
in the range of 0.3--10\,keV with grades 0--2 are included, totalling $\sim$1400 source counts for further analysis. 

For \emph{XMM}-Newton, we perform the data reduction
using Science Analysis Software (SAS version 16.0) and the latest calibration files. In this study,
we only analyze the PN data (Large Window mode), and we process the data in the standard way using the SAS command $epproc$. We  generate a Good Time Interval (GTI) file with the task $tabgtigen$
and filter events with the option  ``RATE$<$=0.4''. We also filter out artifacts from the calibrated and concatenate the dataset with the events screening criteria ``FLAG==0''. We extract source photons from a circular aperture
of 20$''$ centered at the nominal X-ray position (R.A., Decl.) = (11$^{\rm h} 19^{\rm m} 14.26^{\rm s}$,-61$^{\circ}27'49.3''$), and we select the data in 0.15--12 keV energy band. We  obtain   $\sim$26000  counts for Aug.~6
and Aug.~15, $\sim$20000 counts for Aug.~30, and $\sim$5890 counts for Dec. 13, respectively, from the source region.

For \emph{NuSTAR} data analysis, we use the script $nuproducts$ provided by HEASoft,  which runs automatically in sequence all the tasks for the standard data reduction, and we use
the data taken with the Focal Plane Modules A and B (FPM A/B). We generate  source and background  extraction region files using DS9 by choosing a circular region of $\sim 50''$ radius centered on the source and
the background region close to the source, respectively. All required event files and spectral files in this study are obtained  with the task $nuproducts$ and we obtained $\sim$20500 counts for Aug.~6, $\sim$16000 for Aug.~15, $\sim$10500 counts for Aug.~30, and $\sim$2400 counts for Dec.~13. 

To perform the phase-resolved spectroscopy, we define an on-pulse and off-pulse phase for each \emph{XMM}-Newton and \emph{NuSTAR} observation,
as shown in Figure~\ref{lcxmm}. For the \emph{XMM}-Newton data, we extract the events of on/off pulse phase using the phase information obtained from the {$efold$} command.

For the \emph{NuSTAR} data, we create Good Time Interval ($GTI$) files
for on-pulse and off-pulse phases that are identical with the phase-interval for \emph{XMM}-Newton,
and apply the $GTI$ files to $nuproduct$ task with the ``usrgtifile''  parameter.  We group the channels so as to achieve a signal-to-noise
ratio $S/N= 3$ in each energy bin for \emph{XMM}-Newton PN data and at least 30 counts per bin for \emph{NuSTAR} FPM A/B
data. We carry out the spectral analysis with the XSPEC version 12.9.1.

\subsection{Fermi data}
\label{sec:fermi_data}

We use {\it Fermi} Large Area Telescope (LAT)\citep{Atwood2009} Pass 8 (P8R2)\citep{Atwood2013} data with the source centered at RA=$11^{\rm h}19^{\rm m}14\fs{3}$, Decl.=$-61^{\circ}27'49\farcs{5}$ (J2000), the radio coordinates of the pulsar \citep{SK2008}, with an uncertainty of  $0\farcs{3}$.  
We selected ``Source'' class events (evclass= 128) and collected photons from the front and back sections of the tracker (i.e., evttype = 3).
The data quality is further constrained by restriction to instrument good time intervals (i.e., DATA\_QUAL $>$ 0), and we also remove events with zenith angles larger than $90^{\circ}$ to reduce gamma-ray contamination from the earth's limb.

In our spectral analysis, we construct a background emission model, including the Galactic diffuse emission (gll\_iem\_v06) and the isotropic diffuse emission (iso\_P8R2\_SOURCE\_V6\_v06) circulated by the {\it Fermi} Science Support Center, and all 3FGL catalog sources (Acero et al. 2015) within $10^{\rm o}$ of the center of the region of interest.
The phase-averaged spectrum was obtained using the {\it Fermi} Science Tool "$gtlike$" to perform a binned likelihood analysis of data in the energy range 0.1--300\,GeV and within a $20^{\circ}$ radius region of interest (ROI) centered on the pulsar position.

In order to investigate the change of the spectra, we generate the phase-averaged spectra for three epochs determined by the spin-down behavior and GeV flux evolution: pre-outburst epoch (MJD~57023 to MJD~57570), outburst/relaxation epoch (MJD~57570 to MJD~57815) and post-relaxation epoch (MJD~57815 to MJD~58482). 

For the timing analysis, we consider photons within a 2-degree aperture, which contains most of the significant source photons.
We use photon weighting to increase the sensitivity of pulsation statistics \citep{Kerr2011}.  
Using the results of the likelihood analysis, we used the {\it gtdiffrsp} and {\it gtsrcprob} tools to assign a probability to each photon that it originated from the pulsar.  Assigning phases to photons is accomplished with the \textit{Fermi} plugin for \textsc{Tempo2}. 
We obtain 39,045 counts in 0.1--300\,GeV between MJD 57870--58380, while the total weighted photon count is only 1131.2, indicating the aperture is dominated by background.
All the photon arrival times were corrected to the barycentric dynamical time (TDB) with the JPL DE405 solar system ephemeris with the task {\it gtbary}.  The high-order polynomial terms of the timing ephemerides mentioned in section~\ref{timing} are used to describe the complicated timing noise.

\section{Results}

\subsection{Phase-resolved spectroscopy of the X-ray data}
\label{xphase}

\begin{figure}
  \epsscale{1}
  \begin{center}
    \includegraphics[scale=0.45,angle=0]{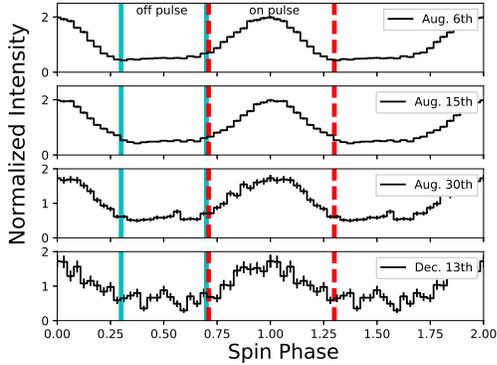}
    \caption{X-ray folded light curves in the 0.5--10.0\,keV energy band from \emph{XMM}-Newton on 2016 August 6, 15, 30 and December 13 from upper to bottom.
      The  vertical green solid-lines and red dashed-lines define  the off-pulse phase and  on-pulse phase, respectively. }
    \label{lcxmm}
  \end{center}
\end{figure}

\begin{figure}
  \epsscale{1}
  \begin{center}
   \includegraphics[scale=0.45,angle=0]{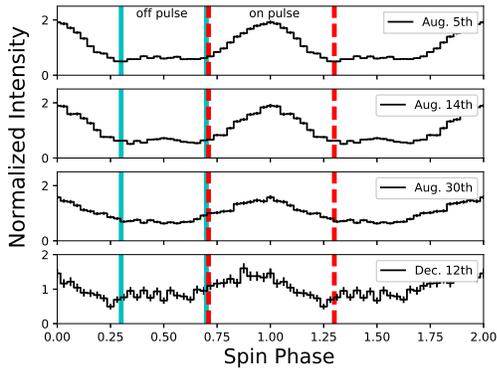}
   \caption{X-ray folded light curves in the 3.0--78.0\,keV energy band from \emph{NuSTAR} on 2016 August 5, 14, 30 and December 12 from upper to bottom.
    The  vertical green solid-lines and red dashed-lines define  the off-pulse phase and  on-pulse phase, respectively}
\label{lcnustar}
\end{center}
 \end{figure}

Figures~\ref{lcxmm} and~\ref{lcnustar} show the folded light curves for \emph{XMM}-Newton and \emph{NuSTAR} observations, respectively, produced as described in \S\ref{xray_analysis}. In each panel, we indicate the definition of the on-pulse and off-pulse phases for phase-resolved spectroscopy.

\begin{figure}
   \epsscale{1}
  \includegraphics[scale=0.3,angle=270]{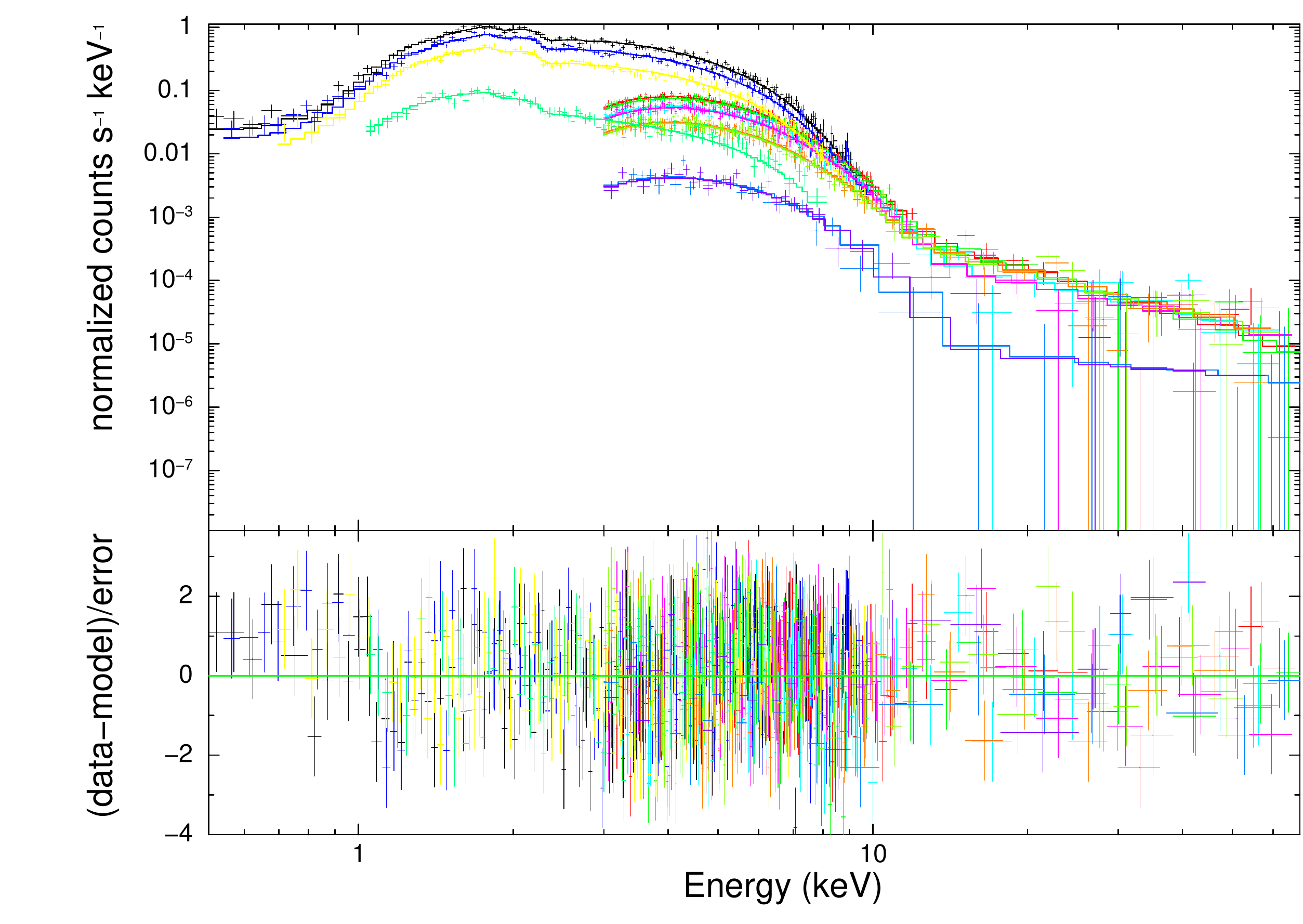}
  \caption{ The joint spectral fits to the on-pulse spectra. In the plots, the spectra extending
    between 0.5-12keV and $>3$keV represent the results for the \emph{XMM}-Newton PN  and \emph{NuSTAR} FPM A/B data, respectively. For each
  instrument, the spectra from top to bottom correspond to the data from 2016 August 6 and December 12/13, respectively.}
\label{allon}
\end{figure}

\begin{figure}
  \epsscale{1}
  \includegraphics[scale=0.3,angle=270]{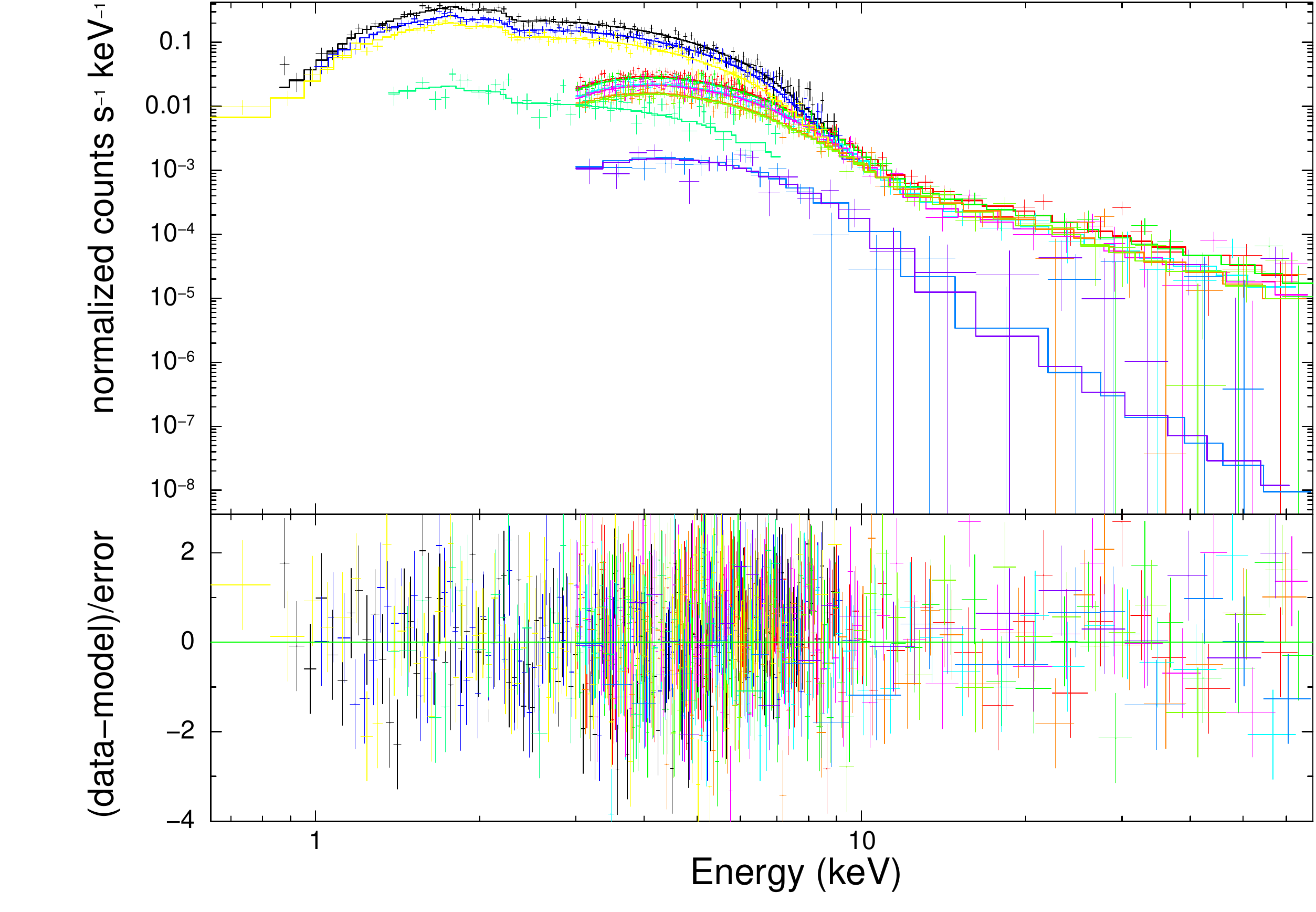}
   \caption{The joint fit to off-pulse spectrum. The colors are as for Figure~\ref{allon}. }
   \label{alloff}
   \end{figure}

Figures~\ref{allon} and~\ref{alloff} show the observed spectra for on-pulse and off-pulse phases, respectively. \textbf{Due to the similar observational time covered by the \emph{XMM}-Newton and \emph{NuSTAR} data, we therefore perform a joint spectral fit of the data obtained from the two instruments. We introduced a constant in the fit to account for the cross-calibration mismatch and modeled photoelectric absorption in the ISM using Wisconsin cross-sections \citep{MC1983}. We fixed the constant factor of \emph{XMM}-Newton at unity, and obtained $\sim1.11$ for \emph{NuSTAR} FPM A/B through the fitting. We use two blackbody emission components plus a power-law with a photon index $(\Gamma)$ and the linked absorption of $N_H = 1.45^{+0.1}_{-0.1} \times 10^{22}$ $cm^{-2}$  to fit all the spectra simultaneously.}  
Table~\ref{epheme1} summarizes the best-fit parameters for on-pulse and off-pulse spectra, respectively. The temperatures and radii of the 2BB components are consistent with the previous study \citep{Lin2018}, which fit the phase-resolved spectra of the \emph{XMM}-Newton data with the 2BB model.  As Table~\ref{epheme1} shows, we find that the power-law component is required for both on- and off-pulse intervals for August data, although the 2BB model sufficiently describes the spectrum of the on-pulse phase on August 15. This is likely because the blackbody emission is $\sim$2 orders of magnitude brighter than the power-law component, and the contribution of the power-law component is not significant on August 15. In December, the 2BB model can provide an acceptable fit to both the on- and off-pulse spectra, and the negative power-law component is poorly constrained and therefore unreliable. Following the evolution of the observed flux in Table~\ref{epheme1}, the flux level of the power-law component decreased below the detector sensitivity by 2016 December.

 The pulse profiles in the 3--78\,keV band (Figure~\ref{lcnustar}) are dominated by the thermal components, but they may not
  represent the pulsation above 10\,keV, where the power-law component dominates. In fact,
  no significant pulsed signal was detected above 10\,keV in August data \citep{Lin2018}.
  
\begin{deluxetable*}{c|cccc}
  \tablecaption{Best-fit spectral parameters obtained from the fit to the phase-resolved spectra of the on-pulse phase and off-pulse phases.
    \label{epheme1}}
  \tablewidth{0pt}
  \tablehead{   & Aug. 5/6   & Aug. 14/15  & Aug. 30  & Dec. 12/13 }
  \multicolumn{5}{c}{On-pulse} \\
  $N_{H}\tablenotemark{a}$($cm^{-2})$ &\multicolumn{4}{c}{1.45$^{+0.1}_{-0.1}\times 10^{22}$} \\
  kT1(KeV) &$0.31^{+0.03}_{-0.02}$&$0.35^{+0.03}_{-0.02}$&$0.35^{+0.04}_{-0.03}$&$0.33^{+0.02}_{-0.03}$ \\
  R1(km) &$5.76^{+1.4}_{-1.47}$&$4.28^{+1.6}_{-0.86}$&$3.7^{+0.93}_{-1.05}$&$2.28^{+0.42}_{-0.33}$ \\
  F$_{BB1}\tablenotemark{b}$&0.45$^{+0.6}_{-0.28}$&0.39$^{+0.4}_{-0.28}$&0.27$^{+0.37}_{-0.12}$&0.088$^{+0.08}_{-0.05}$   \\
  kT2(KeV) &$1.0^{+0.01}_{-0.01}$&$1.02^{+0.01}_{-0.01}$&$1.02^{+0.02}_{-0.02}$&$1.08^{+0.02}_{-0.02}$  \\
  R2(km) &$1.23^{+0.01}_{-0.01}$&$0.98^{+0.03}_{-0.03}$&$0.7^{+0.07}_{-0.07}$&$0.24^{+0.02}_{-0.02}$  \\
  F$_{BB2}\tablenotemark{b}$&2.1$^{+0.35}_{-0.09}$&1.58$^{+0.16}_{-0.13}$&0.98$^{+0.07}_{-0.19}$&0.12$^{+0.05}_{-0.03}$ \\
  $\Gamma \tablenotemark{c}$&$0.84^{+0.2}_{-0.39}$&$0.06^{+0.2}_{-0.2}$&$0.48^{+0.2}_{-0.2}$&$-1.01^{+0.9}_{-0.95}$\\
  F$_{po}\tablenotemark{b}$&0.027$^{+0.022}_{-0.012}$&0.006$^{+0.006}_{-0.003}$&0.014$^{+0.01}_{-0.006}$&1.15E-4$^{+5\rm{E}-6}_{-7\rm{E}-6}$\\
     \multicolumn{5}{c}{Off-pulse} \\
     kT1(KeV) &0.35$^{+0.06}_{-0.04}$&0.32$^{+0.05}_{-0.04}$&0.32$^{+0.06}_{-0.04}$&0.37$^{+0.001}_{-0.001}$ \\
     R1(km) &2.86$^{+1.5}_{-0.7}$&3.07$^{+1.8}_{-1.1}$&2.65$^{+1.7}_{-1.1}$&0.77$^{+0.77}_{-0.8}$ \\
     F$_{BB1}\tablenotemark{b}$&0.18$^{+0.7}_{-0.15}$&0.13$^{+0.5}_{-0.1}$&0.11$^{+0.58}_{-0.08}$&0.016$^{+0.2}_{-0.016}$ \\
     kT2(KeV) &1.05$^{+0.02}_{-0.02}$&1.06$^{+0.02}_{-0.02}$&1.03$^{+0.03}_{-0.03}$&1.2$^{+0.1}_{-0.1}$  \\
     R2(km) &0.65$^{+0.2}_{-0.3}$&0.55$^{+0.33}_{-0.22}$&0.48$^{+0.18}_{-0.2}$&0.11$^{+0.02}_{-0.02}$  \\
     F$_{BB2}\tablenotemark{b}$&0.78$^{+0.12}_{-0.18}$&0.59$^{+0.1}_{-0.09}$&0.44$^{+0.1}_{-0.1}$&0.04$^{+0.04}_{-0.02}$ \\
     $\Gamma \tablenotemark{c}$&0.51$^{+0.3}_{-0.2}$&0.53$^{+0.43}_{-0.37}$&0.89$^{+0.42}_{-0.37}$&$3.11^{+0.92}_{-1.992}$ \\
     F$_{po}\tablenotemark{b}$&0.03$^{+0.011}_{-0.008}$&0.02$^{+0.006}_{-0.018}$&0.03$^{+0.015}_{-0.015}$&0.042$^{+0.03}_{-0.0001}$ \\
      $\chi_{\nu}^2$/{D.O.F.}&\multicolumn{4}{c}{{1.15/2478}}
     \tablenotetext{a}{The  absorption column density.}
     \tablenotetext{b} { The unabsorbed flux is measured in 0.5-10.0 keV and recorded in units of 10$^{-11}$ erg cm$^{-2}$ s$^{-1}$.}
     \tablenotetext{c}{Photon index of the power-law model.}
   \end{deluxetable*}

 \subsection{Comparison between radio and X-ray pulse positions}
\label{peakcom}
 \begin{deluxetable*}{c|cccccc}
     \tablecaption{Ephemeris of radio and X-ray
       \label{epheme3}}
     \tablewidth{0pt}
     \tablehead{  &radio/$t_0$ &X-ray& Instrument & $f_0$ &$f_1$ \\
     & (MJD) & (MJD) & & (Hz) & (${\rm Hz\cdot s^{-1}}$) } \\
     Aug. 9 & 57609.19 &57609.16-57609.83 & \emph{Swift} &2.439814&--\\
     Aug. 15 &57615.07 &57614.50-57615.50 &\emph{NuSTAR} &2.4397973&--\\
     Aug. 30 &57630.17&57630.12-57630.52 &\emph{XMM} &2.4397218&--\\
     Dec. 10-13  &57732.72&57735.38-57735.94 &\emph{XMM} &2.43918&-2.507E-11
\label{ephe}
   \end{deluxetable*}

\begin{figure}
\begin{center}
  \includegraphics[scale=0.46]{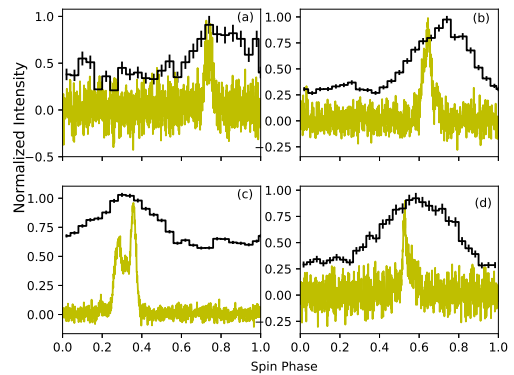}
  \caption{Radio (yellow lines) and X-ray (black histograms) pulse profiles of PSR~J1119$-$6127
    on 2016 (a) August~6, (b) August~15, (c) August~30 and (d) December~10 for radio and December~12/13
    for X-ray. The X-ray folded light curves  are obtained from \emph{Swift} (a), \emph{NuSTAR} (b) and \emph{XMM}-Newton (c and d) observations. We define the phase zero ($t_0$)
  at the epoch of the radio observation (second column in Table~\ref{ephe}).}
 \label{rxlc}
 \end{center}
\end{figure}

Prior to the 2016 outburst, the X-ray pulse peak was aligned with the radio pulse peak \citep{Ng2012}. In this study, therefore, we compare radio pulse profiles reported in \citet{Dai2018} with X-ray pulse profiles taken
in 2016 August and December. Here we used radio observations with the Parkes telescope taken 
on August 9, 15, 30 and December 10, all of which have an observing bandwidth of 
256\,MHz centred at 1369\,MHz. Details of the observations and data reduction can 
be found in \citet{Dai2018}.
Owing to a rapid change in the spin-down rate, it is not practical to create a single global ephemeris
spanning from August to December. Instead, we use a local timing solution determined from each X-ray observation according to the maximum H-test value yielded in a periodicity search.  In our analysis, the radio observations are each spanned by a single X-ray observation at the August 6, 15, and 30 epochs, so that the spin frequency $f_0$ obtained from the X-ray observation can be used to fold the corresponding radio data.  However, in December, the radio and X-ray observations are separated by about two days, necessitating the application of a $f_1$ component in the timing solution used to fold the radio data.  Table~\ref{ephe} summarizes the local ephemerides used in this study, where we define the epoch/phase zero ($t_0$) near the epoch of the radio observation (second column in Table~\ref{ephe}). 

To compare with the radio and X-ray peak positions in 2016 December, we generate a reference $f_0$ and $f_1$ from the information of the local spin frequency determined by the radio observations in 2016 December and 2017 January, assuming that the spin-down rate in 2016 December is stable as reported by \citet{Dai2018}.
To correctly align the radio data, for each radio observation we created an artificial pulse time-of-arrival (TOA) corresponding to the reference epoch in Table~\ref{ephe}, and we likewise used a model of the radio pulse profile (template) to measure the radio TOA at the observatory.  By measuring the offset between these TOAs with \textsc{Tempo2} \citep{hem06}, we obtained the correct value by which to rotate the radio profile.  In this way, all effects of dispersion and light travel time between the barycenter and the Parkes telescope are accounted for.  On the other hand, the barycentered X-ray timestamps are simply folded by the ephemerides in Table~\ref{ephe} such that the epoch represents pulse phase 0.

Figure~\ref{rxlc} compares the pulse profiles of the X-ray and radio emission. 
We find there is no significant evolution in X-ray and radio peak phasing following the X-ray outburst. Moreover, the X-ray light curve consists of one single and broad peak with which the radio peak is roughly aligned, similar to the pre-outburst configuration \citep{Ng2012}.
Even so, the observed X-ray flux remained an order of magnitude higher than that preceding the outburst (Figure~\ref{fitlumi}).

\subsection{\emph{Fermi}-LAT long term light curve and spectra}
\label{gevspec}

\begin{figure}[h]
  \includegraphics[scale=0.4]{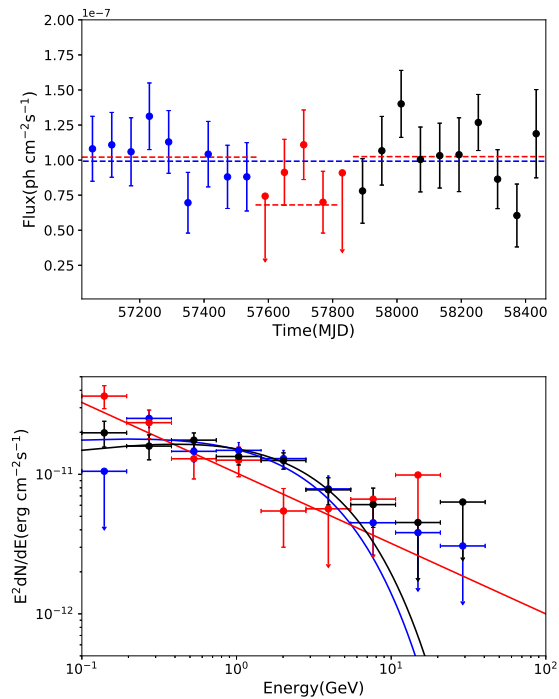}
  \caption{Evolution of the $>0.1$ GeV flux of PSR J1119-6127 (upper panel). The light curve is generated with $\Gamma_2$=1, the blue dashed line shows the time-averaged flux and the red horizontal lines indicate the averaged fluxes for three epochs. Phase-averaged spectra for three epochs(lower panel). Blue: pre-outburst (MJD~57023 to MJD~57570). Red: relaxation (MJD~57570 to MJD~57815).Black: post-relaxation (MJD~57815 to MJD~58482).}
  \label{gall}
\end{figure}

The first panel of Figure~\ref{gall} shows the light curve ($ >$0.1\,GeV)  between MJD~57023 and 58482 obtained with the standard binned maximum likelihood analysis.
We divide the entire time range into 60 day time bins, and we measure an average photon flux of $\sim 9.5\times 10^{-8}$cts cm$^{-2}$ s$^{-1}$.
 The time-averaged photon fluxes for three epochs are
$F= (1.0\pm 0.5)\times 10^{-7}$\,cts cm$^{-2}$ s$^{-1}$ for the pre-outburst epoch, $F= (0.7\pm 0.2)\times 10^{-7}$\,cts cm$^{-2}$ s$^{-1}$ for outburst/relaxation epoch, and
$F=(1.0\pm 0.6)\times 10^{-7}$\,cts cm$^{-2}$ s$^{-1}$ for the post-relaxation epoch.

The lower panel shows the corresponding gamma-ray spectra for the three epochs  (Figure~\ref{gall}), \textbf{which are modeled by a power-law with {\bf an exponential cut-off}:
\begin{equation} 
  dN/dE=N_{0}(E/E_{0})^{-\Gamma_1}{\rm exp} [-(E/E_{C})^{\Gamma_2}],
\end{equation}
where $dN/dE$ is the differential photon rate per unit energy, time, and area, $N_{0}$ is a normalization constant,
$E_{0}$ is the energy scale factor, $\Gamma_1$ is the spectral power-law index , $E_{C}$ is the cutoff energy, and $\Gamma_2$ is the exponential index.
 We obtain the best fitting parameters ($E_C, \Gamma_1, \Gamma_2$)$\sim$($2.6\pm 0.4$\,GeV, $1.9\pm 0.07$, $0.83\pm 0.02$) for the pre-outburst epoch, ($2.2\pm0.3$\,GeV, $2.5\pm0.12$, 0.005$\pm0.001$) for the outburst/relaxation epoch 
 and ($2.6\pm 0.25$\,GeV, $1.92\pm 0.09$, $0.8\pm0.02$) for the post-relaxation state (see Table~\ref{fermiepoch}). 
 We confirm in Figure~\ref{gall} that the phase-averaged
 spectrum in the post-relaxation epoch is consistent with that of pre-outburst epoch. For 
 the outburst/relaxation state, a very small 
 exponential index ($\Gamma_2\sim 0.005$) indicates that a power law function is enough to describe the observed spectra.  We also fit the spectra with a pure  exponential 
 cut-off function ($\Gamma_2=1$) and obtained the fitting parameters of
  ($E_C, \Gamma_1$)=(3.1$\pm 0.45$\,GeV, $1.82\pm0.06$) for the pre-outburst epoch, (2.3$\pm 0.23$\,GeV, $2.16\pm0.21$) for the outburst/relaxation epoch and (3.1$\pm 0.35$\,GeV, $1.85\pm0.09$) for the post-relaxation epoch, respectively. These spectral analyses show the GeV emission during the outburst/relaxation state is significantly softer than ``steady spectra" obtained in pre-outburst and post relaxation states. 
 }
 
  \begin{deluxetable*}{c|ccc}
  \tablecaption{Parameters of spectral fitting of three epochs
    \label{fermiepoch}}
     \tablewidth{0pt}
    \tablehead{Epochs &Pre-outburst &Outburst/Relaxation & Post-relaxation}
      MJD&57023-57570&57570-57815&57815-58482 \\
    Flux($10^{-7}$ cts cm$^{-2}$ s$^{-1}$)&1.02$\pm$ 0.49&0.8$\pm$0.25&1.02$\pm$0.50\\
    Cutoff Energy (GeV)&$2.6\pm 0.4$&$2.2\pm0.3$&$2.6\pm 0.25$\\
    Photon index ($\Gamma_1$) & $1.9\pm 0.07$&$2.5\pm0.12$&$1.92\pm 0.09$\\ 
Exponential index($\Gamma_2$) & 0.83$\pm$ 0.02 &0.005$\pm0.001$ &$0.8\pm0.02$
 \end{deluxetable*}

\subsection{Gamma-ray pulsation}
\label{timing}

Due to the complicated spin-down rate recovery and flux variability described in \S\ref{sec:fermi_data},  
we are unable to either construct a coherent pulsar timing solution with which to fold gamma-ray data or to recover the gamma-ray pulsations in a ``blind search'' over parameters. 
Instead, we restrict our pulsation search and characterization to the ``post-relaxation'' period following MJD~57815.
Using Parkes observations and the resulting pulse times-of-arrival (TOAs), we are able to adequately model the pulsar spin evolution with the ephemeris tabulated in the right column of Table \ref{ephemeris}, whose parameters were optimized using \textsc{Tempo2}.  Additionally, we used pre-outburst radio observations to build a long-term ephemeris (left column).

\begin{figure}[h]
  \includegraphics[scale=0.5]{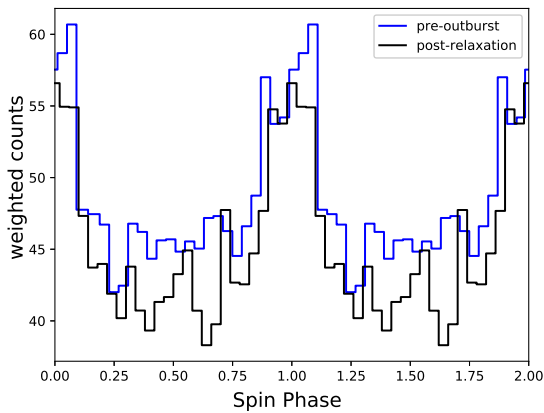} 
\caption{\footnotesize{Gamma-ray pulse profile of PSR~J1119$-$6127 obtained in pre-outburst (MJD 56980-57480) and in post-relaxation (MJD 57872-58379). We obtained the light curve from the weighted photons folded with the ephemerides listed in Table {~\ref{ephemeris}}. The peak positions are shifted to phase zero to compare the pulse structures in two stages, and two cycles of the profile are shown for clarity.
}
\label{fig:gamma_FLC}
}
 \end{figure}

\begin{figure}[h]
  \includegraphics[scale=0.5]{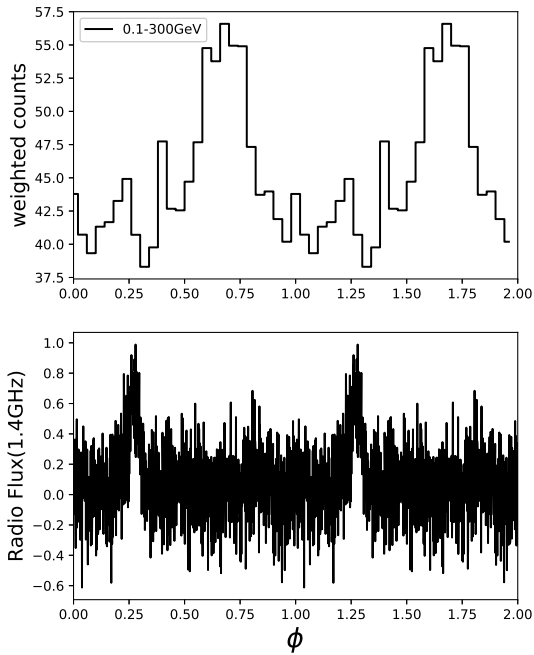} 
\caption{The pulse profiles in $ >$0.1\,GeV gamma-ray (top panel) and radio  (1.4GHz) bands after MJD~57872 (2017 April), which is  $\sim$ 9 months after the X-ray outburst/glitch. The phase zero in each panel refers to MJD~57935.25.
}
\label{fig:radioplusflc}
 \end{figure}

{\bf Using these ephemerides, we folded the gamma-ray data using photon weights \citep{Kerr2011} to enhance the pulsed signal. Fig.~\ref{fig:gamma_FLC} presents
 the resulting 0.1--300\,GeV pulse profiles of PSR~J1119$-$6127 and compares 
the pulse profiled in pre-outburst and post-relaxation stages. No significant 
change of the pulse profile was found.} Fig.~\ref{fig:radioplusflc} compares the pulse profiles between 
 gamma-ray and radio emission after the X-ray outburst, MJD 57872---58378. We find that the radio peak  precedes the main gamma-ray peak by $\sim$0.4 cycle. In the previous studies before the X-ray outburst \citep{Parent2011},  it is found that the  measurement of phase shifts between the gamma-ray peak and radio peak is similar to our detection. Since the spin-down rate and 
gamma-ray emission properties  measured after MJD~57800 have recovered to 
those before the X-ray outburst, the reconfiguration of the magnetosphere after X-ray outburst was complete before MJD 57800. The same phase shift measured between the radio peak and gamma-ray peak suggests that a steady magnetospheric structure established after  the 2016 X-ray outburst is similar to that before the outburst event. We also checked the \emph{NICER} data taken during the valid MJD range of the ephemeris; however, we could not obtain a significant pulsation. This is probably owing to the high flux level of the background emission.

\begin{table*}
\begin{center}
\caption[]{Local ephemeris of PSR~J1119$-$6127 derived through TOA analysis through {\it Fermi} archive.\\}\label{ephemeris}
\begin{tabular}{lll}
\hline\hline
\multicolumn{3}{l}{Parameter} \\
\hline
Pulsar name\dotfill & \multicolumn{2}{c}{PSR~J1119$-$6127} \\
Valid MJD range\dotfill &  56254.719---57495.308 & 57872---58378\\
Right ascension, $\alpha$\dotfill & \multicolumn{2}{c}{11:19:14.3} \\
Declination, $\delta$\dotfill &  \multicolumn{2}{c}{-61:27:49.5} \\
Pulse frequency, $\nu$ (s$^{-1}$)\dotfill & 2.444203(5) &2.438745779(1)  \\
First derivative of pulse frequency, $\dot{\nu}$ (s$^{-2}$)\dotfill &$-1.81(7)\times 10^{-11}$ & $-2.46635(3)\times 10^{-11}$\\
Second derivative of pulse frequency, $\ddot{\nu}$ (s$^{-3}$)\dotfill & $-3.7(6)\times 10^{-19}$&$-1.83(2)\times 10^{-20}$ \\
Third derivative of pulse frequency, $\dddot{\nu}$ (s$^{-4}$)\dotfill & $1.9(4)\times 10^{-26}$ &$1.604(4)\times 10^{-26}$ \\
Fourth derivative of pulse frequency, $\nu^{(4)}$ (s$^{-5}$)\dotfill & $-8(2)\times 10^{-34}$&$-6.11(2)\times 10^{-33}$ \\
Fifth derivative of pulse frequency, $\nu^{5)}$ (s$^{-6}$)\dotfill & $2.8(9)\times 10^{-41}$ &$1.639(4)\times 10^{-39}$ \\
Sixth derivative of pulse frequency, $\nu^{(6)}$ (s$^{-7}$)\dotfill & $-7(2)\times 10^{-49}$&$-3.13(2)\times10^{-46}$ \\
Seventh derivative of pulse frequency, $\nu^{(7)}$ (s$^{-8}$)\dotfill & $1.1(6)\times 10^{-56}$&$3.89(8)\times10^{-53}$ \\
Eighth derivative of pulse frequency, $\nu^{(8)}$ (s$^{-9}$)\dotfill & $-9(4)\times 10^{-65}$&$-2.36(9)\times10^{-60}$ \\
Epoch of frequency determination (MJD)\dotfill & 55440  &57935.25 \\
Time system \dotfill & \multicolumn{2}{c}{TDB}\\
RMS timing residual ($\mu$s)\dotfill &1204.41& 28861.388 \\
Number of time of arrivals\dotfill & 42& 39 \\
\hline
\multicolumn{3}{l}{{\footnotesize Note: the numbers in parentheses denote errors in the last digit}} \\

\end{tabular}
\end{center}
\end{table*}

\section{Discussion}

\begin{figure}
  \includegraphics[scale=0.5,angle=0]{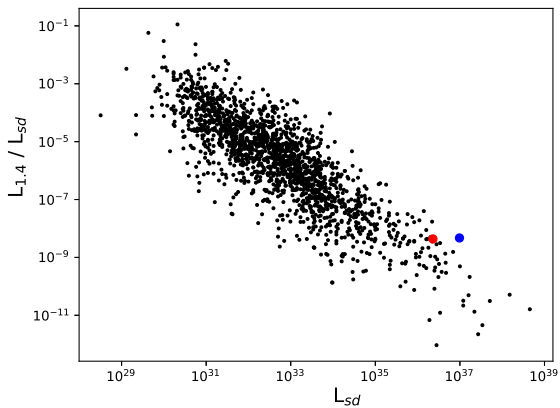}
  \caption{Relation between spin-down power and radio luminosity at 1.4~GHz of the known pulsars. The data come from
    the https://www.atnf.csiro.au/research/pulsar/psrcat/. The red and blue dots are used to denote PSR J1119$-$6127 in the normal stage and soon after its 2016 outburst. }
  \label{lf}
\end{figure}
\label{discuss}
\begin{figure}
\begin{center}
  \includegraphics[scale=0.5]{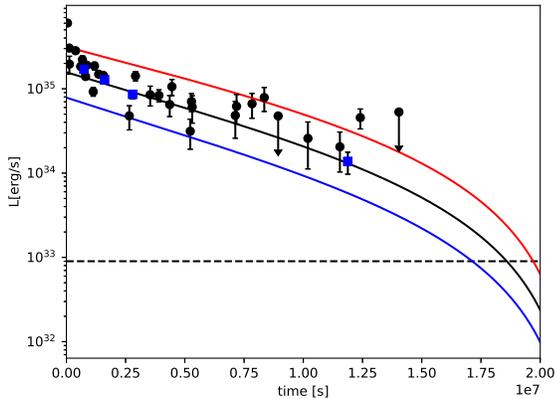}
  \end{center}
\caption {Evolution of the X-ray  luminosity after 2016 outburst, \textbf{The inferred luminosity} by \emph{Swift}(black point) and \emph{XMM}-Newton(blue square) is calculated by assuming the solid angle of 4$\pi$. The solid lines show the model curve predicted by the twisting magnetosphere  \citep{Beloborodov2009},  assuming $\mu=5\times 10^{31}\rm{G \cdot cm^3}$, the initial twist angle of $\phi_0=0.25$ and the initial polar angle of the j-bundle of $u_0=0.4$.  In addition, the electric potential drop is assumed to be $V_e(u)=V'(u_0-u)+V_0$ with $V'=2.5\times 10^9$\,V and $V_0=10^{10}$\,V for red line, $V_0=5\times 10^9$\,V for black line and  $V_0=2.5\times 10^9$\,V for blue line. \bf{The horizontal dash line shows the persistent emission luminosity $L_X\sim 9\times 10^{32}~{\rm erg~s^{-1}}$ }}
\label{fitlumi}
   
\end{figure}

The magnetar XTE~J1810$-$197, with a surface magnetic field of $B_s\sim 3\times 10^{14}$\,G \citep{Ibrahim04, weng15}, underwent an X-ray outburst in 2003 and emitted pulsed radio emission following the outburst.
The evolution of the timing solution, radio emission and X-ray emission properties  of PSR ~J1119$-$6127
after its 2016 outburst are very similar to those of XTE~J1810$-$197, although the recovery time scale ($\tau_{r}$) and released total energy ($E_{tot}$)  are one or two orders of magnitude smaller:
$\tau_{r}\sim 0.5$\,yr and $E_{tot}\sim 10^{42}$\,erg for PSR~J1119$-$6127, and  $\tau_{r}\sim 3$\,yr
and $E_{tot}\sim 10^{43-44}$\,erg for XTE~J1810$-$197 \citep{Camilo2006}.
After the  outburst,  the spin-down rate increased with a time scale of $\sim$30\,d for PSR~J1119$-$6127 and of $\sim$ 2\,yr for  XTE~J1810$-$197, and then recovered toward the steady state values over a time scale $\sim$ 100\,d for PSR~J1119$-$6127 and with $\sim$ 3\,yr for XTE~J1810$-$197.
XTE~J1810$-$197 was known to have radio pulsations \citep{Camilo2006}. The radio emission from the magnetar disappeared in 2008 and reappeared in late 2018 along with an X-ray enhancement (\citealt{Dai19,got19}). This 2018 XTE~J1810$-$197 outburst shows similarities to the 2003 outburst in terms of X-ray flux, spectrum, and pulse properties.  Likewise, the spectra can be fit by two blackbodies plus an additional power-law component to account for emission above 10 keV.  Moreover, \citet{got19} find that the pulse peak of the X-rays lags the radio pulse by $\sim 0.13$ cycles, which is almost exactly as the behavior of the 2003 outburst.  Both XTE~J1810$-$197 and PSR~J1119$-$6127
showed a continuous decrease of the X-ray emission after the outburst.
The similarities between these two pulsars support the identification of the 2016 X-ray outburst of PSR~J1119$-$6127 as magnetar-like activity.

One interesting observational property after the X-ray outburst is the evolution of the radio emission
and spin-down rate  (Figure ~\ref{glc}); (i) the flux evolution
of the radio emission is correlated with the spin-down rate evolution and (ii)
an additional component of the radio emission appeared in 2016 August/September \citep{Majid2017, Dai2018}.
An additional component  was also observed after previous glitches
\citep{WJE2011,Anton2015}. Figure~\ref{lf} plots the efficiency of the 1.4\,GHz radiation ($\eta=L/L_{sd}$)
and the spin-down power for the rotation powered pulsars, for  which the data are taken
from the ATNF pulsar catalog (Manchester et al. 2005) and the radio luminosity is calculated from $L_{1.4}=S_{1400}d^2$ with $d$ being the distance to the source; the red and blue dots represent PSR~J1119$-$6127 ($d=8.4$\,kpc)
in its baseline stage and at the end of 2016 August, respectively.

The evolution of the radio emission and spin-down rate after the X-ray outburst might be a result of the reconfiguration
of the global magnetosphere (current structure and/or dipole moment strength) caused by,  for example, crustal motion
driven by the internal magnetic stress  \citep{Beloborodov2009, Huang2016}.
The appearance of  new radio components after the X-ray outburst may be a result of the change of the structure of
the open magnetic field region. For example, the radio intensity increased by about a factor of 5 in 2016 August after X-ray outburst. \textbf{We suppose this extra component exists} but does not point toward the observer in the normal state. In such a case,  the radio efficiency of this pulsar would be much  higher  than  other pulsars having similar spin-down power unless the distance is less than $d=8.4$\,kpc,
since the apparent efficiency in the normal state is already higher than typical values, as Figure~\ref{lf} shows.
It is also  possible that the extra radio component appearing after the X-ray outburst originates from a new emission
region created after the outburst, and plasma flow on the new open field lines would produce the extra radio emission  after X-ray outburst. The evolution of the spin-down rate and the radio/GeV emission relationship suggest that the re-configuration of the global magnetosphere at most continued for six months after the X-ray outburst and the structure returned to the pre-outburst state by about 2016 December.

Figure~\ref{fitlumi} shows the evolution of the X-ray luminosity (by assuming $d=8.4$kpc) after the X-ray outburst.
The X-ray luminosity gradually decreased after the outburst, and the source was undetected by \emph{Swift} observations in 2017.
In the twisted magnetosphere model \citep{Beloborodov2009}, for example, the timescale of the evolution of the magnetosphere ($t_t$) is
related to the evolution of the current $j$-bundle, and it is estimated as
\begin{eqnarray}
  t_t&=&\frac{\phi_0u\mu}{cR_{NS} V_e}=0.33{\rm yr}\nonumber \\
  &\times&\left(\frac{\phi_0}{0.25}\right)\left(\frac{u}{0.4}\right)
  \left(\frac{\mu}{5\cdot 10^{31}\rm{G~cm^3}}\right)\left(\frac{V_e}{5\cdot10^9{\rm V}}\right)^{-1},
\end{eqnarray}
where $\phi_0\sim B_t/B$ is the twist angle with $B_t$ being the toroidal field, $u=\sin^2 \theta$ with $\theta$ is the polar angle of the current  $j$-bundle,  $R_{NS}=10^6$cm is the neutron star radius and $\mu$ is the magnetic dipole moment. In addition,  $V_e$  is the  voltage to maintain the twist current, which  is expressed by  $I_e\sim \mu c\phi_{0} u^2/2R_{NS}^2$. The induced electric potential may be on the order of $10^9$\,V -$10^{10}$\,V, which can produce GeV electrons that resonantly scatter X-ray to GeV gamma rays that are eventually converted into electron/positron pairs (Beloborodov 2009).
The radiation luminosity, which is $L_t\sim I_eV_e$,  is estimated  to be
\begin{eqnarray}
  L_t &\sim &2.5\times 10^{35}{\rm erg/s} \nonumber \\
  &\times&\left(\frac{\phi_0}{0.25}\right)\left(\frac{u}{0.4}\right)^2
  \left(\frac{\mu}{5\cdot 10^{31}\rm{G~cm^3}}\right)\left(\frac{V_e}{5\cdot10^9{\rm V}}\right).
\end{eqnarray}
Luminosity is roughly proportional to  $(t-t_t)^2$ \citep{Beloborodov2009}. The solid lines in  Figure~\ref{fitlumi} show examples of the evolution of the X-ray luminosity with particular parameters (e.g., twist angle and potential drop), for
which we apply Eq.\,(37) in \citet{Beloborodov2009}  to calculate the temporal evolution of the j-bundle.
This model suggests that the X-ray emission powered by the untwisting  magnetosphere dominates
the persistent emission, $L_X\sim 9\times 10^{32}~{\rm erg~s^{-1}}$ (horizontal  dashed line in Figure~\ref{fitlumi}), until $\sim$0.6\,yr after the outburst.

Although the X-ray spectrum requires two blackbody components, the pulse profile shows a single broad peak after the X-ray outburst, indicating that the emission regions of the two components should be close to each other.  In fact, some young pulsars and millisecond pulsars also show two-blackbody emission from the heated polar cap, in the form of core (high-temperature
and smaller emission region) and rim (low-temperature and wider emission region) components, respectively \citep{Takata2012}. While the incoming
particles heat a part of the stellar surface upon impact, they
also emit high-energy photons toward the stellar surface. Since the local magnetic field line has a curvature, the radiation
heats a  surface area wider than that heated by the bombardment of the incoming particles.  We therefore speculate that two-black body emission of PSR~J1119$-$6127 might also be due to the photon-pair shower, namely, impact of incoming particles (umbra) and illumination by incoming radiation (penumbra).

\textbf{
As presented in sections~\ref{gevspec} and~\ref{timing},  the GeV emission in the relaxation state is probably suppressed due to strong X-ray emission that increases the optical depth of the photon-photon pair-creation process and/or due to a reconfiguration of the magnetosphere. The GeV emission properties (pulse shape, spectrum and phase lag from the radio peak) taken after 2016 December, on the other hand, are consistent with those before the X-ray outburst. These results indicate that the 2016 X-ray outburst and subsequent reconfiguration of the magnetosphere did not permanently change the structure of the GeV emission region. The phase-averaged spectra in the pre-outburst and post-relaxation states can be fitted by the power-law with sub-exponential cut-off ($\Gamma_2\sim 0.8$) model. This sub-exponential cut-off behavior is a common property for most gamma-ray pulsars (Fermi-LAT pulsar catalogue;  \citep{Abdo2013}), and it probably suggests the emission origin from the outer magnetosphere (e.g., \citealt{Aliu2008}; \citealt{Takata2016}). GeV emission from the polar cap accelerator/cascade regions is highly suppressed by the magnetic pair-creation process for the canonical pulsars, and the spectrum of such emission is expected to follow a super-exponential cutoff ($\Gamma_2>1$) in the 0.1-1GeV band.  The observed phase lag of $\sim 0.4$ between the radio and gamma-ray pulse peaks of PSR~J1119$-$6127 also suggests the GeV emission originated from the outer magnetosphere.
The phase lag between the radio  and gamma-ray peaks is also a common feature of the radio-loud gamma-ray pulsars, and the magnitudes distribute between 0-0.5 phase lag with most cases of only around 0.1-0.2 phase lag. The observed phase lag is anti-correlated to the gamma-ray peak separation; smaller phase lag tends to have larger gamma-ray peak separation, and gamma-ray pulsars with a larger phase lag, say 0.3-0.4, tend to have  a single gamma-ray peak \citep{Abdo2013}.   }

\textbf{
A phase lag can be expected if the GeV emission is from the outer magnetosphere and the radio emission is from a region above the polar cap \citep{Romani1996}.  \citet{Pierbattista2016} studied the relationship of phase lag between the gamma-ray and radio peaks and the locations of the GeV and radio emission regions. Their results show that  such a large phase lag for PSR~J1119$-$6127 is difficult to reproduce if the gamma-ray emission comes from the polar cap region. They also demonstrate that the GeV emission from the outer magnetosphere is consistent with the observed relation between the phase lag and gamma-ray peak separation.  }

\textbf{A large lag with a 
single gamma-ray peak of PSR~J1119$-$6127 will be consistent with the GeV emission originated from the outer magnetosphere. For the emission from outer magnetosphere, the line of sight close to the spin equator predicts the double peak GeV light curve, while, with a  smaller or larger viewing angle from the spin axes, say  $\sim 45^{\circ}$ or $135^{\circ}$, it expects a single broad pulse of PSR~J1119$-$6127  \citep{WR2011, Takata2011, Kalapotharakos2014}. Since the GeV emission from  PSR~J1119$-$6127 likely originates from the outer magnetosphere, the suppression of the GeV emission in the relaxation stage suggests that the 2016 magnetar-like outburst affected the structure of the global magnetosphere, and such an influence probably at most continued for about six months. }

 \section{Summary} 
In this study, we have performed a multi-wavelength study for PSR~J1119$-$6127 after its 2016 magnetar-like outburst.
We compared the X-ray and radio pulse profiles measured in 2016 August and December.We found that the X-ray and radio peaks were roughly aligned at different epochs.
From the joint phase-resolved spectrum in different epochs, we found that the observed X-ray spectra of both on-pulse and off-pulse phases in 2016 August are well described by two blackbody components plus a power-law (a photon index $\Gamma=0.5-1$) model.
The power-law component did not show a significant modulation with the spin phase. 

In gamma-ray bands, the GeV emission might be slightly suppressed around X-ray outburst. 
Based on the evolution of the GeV flux and of the spin-down rate, we divided the \emph{Fermi} data into three epochs; pre-outburst epoch, outburst/relaxation epoch in 2016 July-2017 January and post-relaxation epoch after 2017 January.
We found that the GeV spectral characteristics in the post-relaxation epoch are consistent with that of the pre-outburst epoch. Moreover, we confirmed  the gamma-ray pulsation in post-relaxation epoch and the pulse shape is also consistent in terms of pulse profile with that of the pre-outburst stage. The phase difference between the gamma-ray peak and radio peak in post-relaxation stage is $\sim 0.4$, which is consistent with the measurement before the X-ray outburst.

The radio/X-ray emission properties and the spin-down properties of  PSR~J1119$-$6127 after the X-ray outburst  are similar to those of the magnetar XTE~J1810$-$197. The similarities between two pulsars support the identification of the 2016 X-ray
outburst of PSR J1119$-$6127 as magnetar-like activities. The 2016 X-ray outburst probably caused a re-configuration of the global magnetosphere and changed  the structure of the open field line regions. The temporal evolution of the spin-down rate and the radio emission after the X-ray outburst will be related to the evolution of the structure of the open field line regions. The evolution of the X-ray emission will be related to  the heating of the crust  and/or stellar surface.  We reproduce the evolution of the observed X-ray flux with the untwisting magnetosphere model with particular model parameters (e.g., twist angle and potential drop).  The multi-wavelength emission properties suggest that the reconfiguration  of the global magnetosphere continued about a half-year after the X-ray outburst.  The observed relation in the phase difference of the radio/gamma-ray peak positions before and after the X-ray outburst would suggest that the structure of magnetosphere has recovered to a normal state within $\sim 0.6$ years after the outburst.

\acknowledgements
We thank to referee for his/her useful comments and suggestions. We thank to  M. Kerr for his useful discussion and suggestions.  We also appreciate Drs Matthew Baring, D.J. Thompson and P. Bruel for their critical reading to polish our paper. In addition, we thank to N. K. Y. Low for his help to check the timing solution derived in the radio band.
W.H.H. and J.T. are supported by the National Science Foundation of China (NSFC) under 11573010, 11661161010, U1631103 and U1838102. L.C.-C.L. and K.L.L. are supported by the National Research Foundation of Korea (NRFK) through grant 2016R1A5A1013277. C.-P.H. is supported by a GRF grant of the Hong Kong Government under HKU 17300215. X.H. is supported by the National Natural Science Foundation of China (NSFC-U1938103 and NSFC-11673060) and the Light of West China Program of the Chinese Academy of Sciences.  

The \textit{Fermi} LAT Collaboration acknowledges generous ongoing support
from a number of agencies and institutes that have supported both the
development and the operation of the LAT as well as scientific data analysis.
These include the National Aeronautics and Space Administration and the
Department of Energy in the United States, the Commissariat \`a l'Energie Atomique
and the Centre National de la Recherche Scientifique / Institut National de Physique
Nucl\'eaire et de Physique des Particules in France, the Agenzia Spaziale Italiana
and the Istituto Nazionale di Fisica Nucleare in Italy, the Ministry of Education,
Culture, Sports, Science and Technology (MEXT),  High Energy Accelerator Research
Organization (KEK) and Japan Aerospace Exploration Agency (JAXA) in Japan, and
the K.~A.~Wallenberg Foundation, the Swedish Research Council and the
Swedish National Space Board in Sweden.
 
Additional support for science analysis during the operations phase is gratefully
acknowledged from the Istituto Nazionale di Astrofisica in Italy and the Centre
National d'\'Etudes Spatiales in France. This work performed in part under DOE
Contract DE-AC02-76SF00515.
This work made use of data supplied by the LAT data server of the Fermi Science Support Center (FSSC) and the archival data server of NASA's High Energy Astrophysics Science Archive Research Center (HEASARC).

{\it Facilities}: \emph{Parkes, Fermi(LAT), Swift(XRT), XMM(EPIC), NuSTAR.}


\begin{thebibliography}
\expandafter\ifx\csname natexlab\endcsname\relax\def\natexlab#1{#1}\fi
\bibitem[{{Abdo} {et al}.(2013)}]{Abdo2013} 
{Abdo}, A.~A., {Ajello}, M., {Allafort}, A., {Baldini}, L., {et al.} 2013, \apjs, 208, 17
\bibitem[{{Antonopoulou} {et~al.}(2015){Antonopoulou}, {Weltevrede},
  {Espinoza}}]{Anton2015}
{Antonopoulou}, D., {Weltevrede}, P., {Espinoza}, C.~M., {et~al.} 2015, \mnras, 447, 3924
\bibitem[Aliu et al.(2008)]{Aliu2008} Aliu, E., Anderhub, H., Antonelli, L.~A., et al.\ 2008, Science, 322, 1221
\bibitem[{{Archibald} {et~al.}(2016){Archibald}, {Kaspi}, {Tendulkar}}]{Archibald2016}
{Archibald}, R.~F., {Kaspi}, V.~M., {Tendulkar}, S.~P., {et~al.} 2016,
  \apjl, 829, L21
\bibitem[{{Archibald} {et~al.}(2017){Archibald},{Burgay},{Lyutikov},{Kaspi}}]{Archibald2017}
{Archibald}, R. F., {Burgay}, M., {Lyutikov}, M., {Kaspi}, V. M., {et~al.}  2017, ApJL, 849, L20

\bibitem[{{Archibald} {et~al.}(2018){Archibald}, {Kaspi}, {Tendulkar}, \&
  {Scholz}}]{Archibald2018}
{Archibald}, R.~F., {Kaspi}, V.~M., {Tendulkar}, S.~P., {et~al.} 2018, \apj, 869, 180

\bibitem[{{Arons}(1983)}]{Arons83}
{Arons}, J. 1983, \apj, 266, 215
\bibitem[Atwood et al.(2009)]{Atwood2009} Atwood, W.~B., Abdo, A.~A., Ackermann, M., et al.\ 2009, \apj, 697, 1071
\bibitem[Atwood et al.(2013)]{Atwood2013} Atwood, W., Albert, A., Baldini, L., et al.\ 2013, arXiv e-prints, arXiv:1303.3514
\bibitem[{{Beloborodov}(2009)}]{Beloborodov2009}
{Beloborodov}, A.~M. 2009, \apj, 703, 1044

\bibitem[{{Blumer} {et~al.}(2017){Blumer}, {Safi-Harb}, \&
  {McLaughlin}}]{BSM2017}
{Blumer}, H., {Safi-Harb}, S., \& {McLaughlin}, M.~A. 2017, \apjl, 850, L18

\bibitem[{{Burgay} {et~al.}(2016){Burgay}, {Possenti}, {Kerr}, {Esposito}}]{Burgay2016}
{Burgay}, M., {Possenti}, A., {Kerr}, M., {Esposito}, P., {et~al.}2016, The Astronomer's Telegram,
  9366

\bibitem[{{Camilo} {et~al.}(2000){Camilo}, {Kaspi}, {Lyne}}]{Camilo2000}
{Camilo}, F., {Kaspi}, V.~M., {Lyne}, A.~G., et al., 2000, \apj, 541,
  367
  
\bibitem[{{Camilo} {et~al.}(2006){Camilo}, {Ransom}, {Halpern}}]{Camilo2006}
{Camilo}, F., {Ransom}, S.~M., {Halpern}, J.~P., {Reynolds}, J., {Helfand},
  D.~J., {et~al.} 2006, \nat, 442, 892

\bibitem[{{Caswell} {et~al.}(2004){Caswell}, {McClure-Griffiths}, \&
  {Cheung}}]{CMC2004}
{Caswell}, J.~L., {McClure-Griffiths}, N.~M., \& {Cheung}, M.~C.~M. 2004,
  \mnras, 352, 1405

\bibitem[{{Cheng} {et~al.}(1986){Cheng}, {Ho}, \& {Ruderman}}]{CHR86}
{Cheng}, K.~S., {Ho}, C., \& {Ruderman}, M. 1986, \apj, 300, 500

\bibitem[{{Crawford} {et~al.}(2001){Crawford}, {Gaensler}, {Kaspi}}]{Crawford01}
{Crawford}, F., {Gaensler}, B.~M., {Kaspi}, V.~M., {et~al.} 2001, \apj, 554, 152

\bibitem[{Dai {et~al.}(2018)Dai, Johnston, Weltevrede}]{Dai2018}
Dai, S., {Johnston}, S., {Weltevrede}, P.,  {et~al.} 2018, Monthly Notices of the Royal Astronomical Society, 480,
  3584

\bibitem[{Dai {et~al.}(2019)Dai, Lower, Bailes}]{Dai19}
Dai, S., {Johnston}, S., {Weltevrede}, P.,  {et~al.} 2019, The Astrophysical Journal, 874, L14
\bibitem[{De~Jager \& B{\"u}sching(2010)}]{de2010h}
De~Jager, O., \& B{\"u}sching, I. 2010, Astronomy \& Astrophysics, 517, L9

\bibitem[{{Dib} {et~al.}(2008){Dib}, {Kaspi}, \& {Gavriil}}]{DKG2008}
{Dib}, R., {Kaspi}, V.~M., \& {Gavriil}, F.~P. 2008, \apj, 673, 1044

\bibitem[{{Enoto} {et~al.}(2017){Enoto}, {Shibata}, {Kitaguchi}}]{Enoto2017}
{Enoto}, T., {Shibata}, S., {Kitaguchi}, T., {et~al.} 2017, \apjs, 231, 8

\bibitem[{{Gonzalez} \& {Safi-Harb}(2003)}]{GS03}
{Gonzalez}, M., \& {Safi-Harb}, S. 2003, \apjl, 591, L143


\bibitem[{Gotthelf} {et~al.}(2019)]{got19} {Gotthelf}, {E.~V.}, {Halpern}, {J.~P.}, {Alford}, J.~A.~J., {et al}. 2019, arXiv:1902.08358 
\bibitem[{{G{\"o}{\u g}{\"u}{\c s}} {et~al.}(2016){G{\"o}{\u g}{\"u}{\c s}},
  {Lin}, {Kaneko}, {Kouveliotou}, {Watts}, {Chakraborty}, {Alpar},
  {Huppenkothen}, {Roberts}, {Younes}, \& {van der Horst}}]{Gogus2016}
{G{\"o}{\u g}{\"u}{\c s}}, E., {Lin}, L.,  {Kaneko}, Y., {et~al.} 2016, \apjl, 829, L25
\bibitem[{{Hobbs}, {Edwards} \& {Manchester}(2006){Hobbs}, {Edwards}, \&
   {Manchester}}]{hem06} {Hobbs} G.~B., {Edwards} R.~T., {Manchester} R.~N., 2006, MNRAS, 369, 655
\bibitem[Hotan et al.(2004)]{2004PASA...21..302H} Hotan, A.~W., van Straten, W., \& Manchester, R.~N.\ 2004, Publications of Astronomical Society of Australia, 21, 302

\bibitem[{Huang {et~al.}(2016)Huang, Yu, \& Tong}]{Huang2016}
Huang, L., Yu, C., \& Tong, H. 2016, The Astrophysical Journal, 827, 80

\bibitem[{{Ibrahim} {et~al.}(2004){Ibrahim}, {Markwardt}, {Swank}, {Ransom},
  {Roberts}, {Kaspi}, {Woods}, {Safi-Harb}, {Balman}, {Parke}, {Kouveliotou},
  {Hurley}, \& {Cline}}]{Ibrahim04}
{Ibrahim}, A.~I., {Markwardt}, C.~B., {Swank}, J.~H., {et~al.} 2004, \apjl, 609, L21

\bibitem[Johnston \& Weisberg(2006)]{2006MNRAS.368.1856J} Johnston, S., \& Weisberg, J.~M.\ 2006, \mnras, 368, 1856 
\bibitem[Kalapotharakos et al.(2014)]{Kalapotharakos2014} Kalapotharakos, C., Harding, A.~K., \& Kazanas, D.\ 2014, \apj, 793, 97

\bibitem[{{Kaspi} {et~al.}(2003){Kaspi}, {Gavriil}, {Woods}, {Jensen},
  {Roberts}, \& {Chakrabarty}}]{Kaspi03}
{Kaspi}, V.~M., {Gavriil}, F.~P., {Woods}, P.~M., {Jensen}, J.~B., {Roberts},
  M.~S.~E., \& {Chakrabarty}, D.,{et~al.}  2003, \apjl, 588, L93
\bibitem[{{Kaspi} \& {Beloborodov}(2017)}]{Kaspi17}
{Kaspi}, V.~M., \& {Beloborodov}, A.~M. 2017, \araa, 55, 261
\bibitem[{{Kennea} {et~al.}(2016){Kennea}, {Lien}, {Marshall}, {Palmer},
  {Roegiers}, \& {Sbarufatti}}]{Kennea2016}
{Kennea}, J.~A., {Lien}, A.~Y., {Marshall}, F.~E., {et~al.} 2016, GRB Coordinates Network, Circular
  Service, No.~19735, \#1 (2016), 19735

\bibitem[{{Kerr}(2011)}]{Kerr2011}
{Kerr}, M. 2011, \apj, 732, 38

\bibitem[{{Lin} {et~al.}(2018){Lin}, {Wang}, {Li}}]{Lin2018}
{Lin}, L.~C.-C., {Wang}, H.-H., {Li}, K.-L., {et~al.} 2018, \apj, 866, 6

\bibitem[{{Livingstone} {et~al.}(2010){Livingstone}, {Kaspi}, \&
  {Gavriil}}]{Livingstone2010}
{Livingstone}, M.~A., {Kaspi}, V.~M., \& {Gavriil}, F.~P. 2010, \apj, 710, 1710

\bibitem[{{Livingstone} {et~al.}(2009){Livingstone}, {Ransom}, {Camilo}}]{Livingstone2009}
{Livingstone}, M.~A., {Ransom}, S.~M., {Camilo}, F., {et~al.} 2009, \apj, 706, 1163
\bibitem[{{Lyne} {et~al.}(2018){Lyne}, {Levin}, {Stappers}, {Mickaliger},
  {Desvignes}, \& {Kramer}}]{lyne18}
{Lyne}, A., {Levin}, L., {Stappers}, B., {et~al.} 2018, The Astronomer's Telegram, 12284

\bibitem[{{Majid} {et~al.}(2017){Majid}, {Pearlman}, {Dobreva}, {Horiuchi},
  {Kocz}, {Lippuner}, \& {Prince}}]{Majid2017}
{Majid}, W.~A., {Pearlman}, A.~B., {Dobreva}, T., {et~al.} 2017, \apjl, 834, L2

\bibitem[Manchester et al.(2005)]{2005AJ....129.1993M} Manchester, R.~N., Hobbs, G.~B., Teoh, A., \& Hobbs, M.\ 2005, \aj, 129, 1993
\bibitem[{{Mihara} {et~al.}(2018){Mihara}, {Negoro}, {Kawai}, {Nakajima},
  {Maruyama}, {Sakamaki}, {Aoki}, {Kobayashi}, {Nakahira}, {Yatabe}, {Takao},
  {Matsuoka}, {Sakamoto}, {Serino}, {Sugita}, {Kawakubo}, {Hashimoto},
  {Yoshida}, {Sugizaki}, {Tachibana}, {Morita}, {Oeda}, {Shiraishi}, {Ueno},
  {Tomida}, {Ishikawa}, {Sugawara}, {Isobe}, {Shimomukai}, {Midooka}, {Ueda},
  {Tanimoto}, {Morita}, {Yamada}, {Ogawa}, {Tsuboi}, {Iwakiri}, {Sasaki},
  {Kawai}, {Sato}, {Tsunemi}, {Yoneyama}, {Asakura}, {Ide}, {Yamauchi},
  {Hidaka}, {Iwahori}, {Kurihara}, {Kawamuro}, {Yamaoka}, \&
  {Shidatsu}}]{mihara18}
{Mihara}, T., {Negoro}, H., {Kawai}, N., {et~al.} 2018, The Astronomer's Telegram, 12291

\bibitem[{{Morrison}\&{McCammon}(1983)}] {MC1983}
{Morrison}, R., \& {McCammon}, D. 1983, \apj, 270, 119
\bibitem[{{Ng} {et~al.}(2012){Ng}, {Kaspi}, {Ho}, {Weltevrede}, {Bogdanov},
  {Shannon}, \& {Gonzalez}}]{Ng2012}
{Ng}, C.-Y., {Kaspi}, V.~M., {Ho}, W.~C.~G., {et~al.} 2012, \apj, 761, 65

\bibitem[{{Ng} {et~al.}(2011){Ng}, {Kaspi}, {Dib}, {Olausen}, {Scholz},
  {G{\"u}ver}, {{\"O}zel}, {Gavriil}, \& {Woods}}]{Ng2011}
{Ng}, C.-Y., {Kaspi}, V.~M., {Dib}, R., {et~al.} 2011, \apj, 729, 131

\bibitem[{{Parent} {et~al.}(2011){Parent}, {Kerr}, {den Hartog}, {Baring},
  {DeCesar}, {Espinoza}, {Gotthelf}, {Harding}, {Johnston}, {Kaspi},
  {Livingstone}, {Romani}, {Stappers}, {Watters}, {Weltevrede}, {Abdo},
  {Burgay}, {Camilo}, {Craig}, {Freire}, {Giordano}, {Guillemot}, {Hobbs},
  {Keith}, {Kramer}, {Lyne}, {Manchester}, {Noutsos}, {Possenti}, \&
  {Smith}}]{Parent2011}
{Parent}, D., {Kerr}, M., {den Hartog}, P.~R., {et~al.} 2011, \apj, 743, 170

\bibitem[Petroff et al.(2013)]{2013MNRAS.435.1610P} Petroff, E., Keith, M.~J., Johnston, S., {et~al.} \ 2013, \mnras, 435, 1610 

\bibitem[{{Pierbattista} et al.(2016)}]{Pierbattista2016} 
{Pierbattista}, M., {Harding}, A.~K., {Gonthier}, P.~L., {Grenier}, I.~A. 2016, \aap, 588, A137

\bibitem[{Pons \& Rea(2012)}]{Pons2012}
Pons, J.~A., \& Rea, N. 2012, The Astrophysical Journal Letters, 750, L6

\bibitem[{{Ray} {et~al.}(2011){Ray}, {Kerr}, {Parent}, {Abdo}, {Guillemot},
  {Ransom}, {Rea}, {Wolff}, {Makeev}, {Roberts}, {Camilo}, {Dormody}, {Freire},
  {Grove}, {Gwon}, {Harding}, {Johnston}, {Keith}, {Kramer}, {Michelson},
  {Romani}, {Saz Parkinson}, {Thompson}, {Weltevrede}, {Wood}, \&
  {Ziegler}}]{Ray2011}
{Ray}, P.~S., {Kerr}, M., {Parent}, D., {et~al.} 2011, \apjs, 194, 17

\bibitem[{{Romani}(1996)}]{Romani1996} {Romani}, R.~W.\ 1996, \apj, 470, 469

\bibitem[{{Safi-Harb} \& {Kumar}(2008)}]{SK2008}
{Safi-Harb}, S., \& {Kumar}, H.~S. 2008, \apj, 684, 532
\bibitem[Takata et al.(2011)]{Takata2011} Takata, J., Wang, Y., \& Cheng, K.~S.\ 2011, \mnras, 415, 1827
\bibitem[{{Takata} {et~al.}(2012){Takata}, {Cheng}, \& {Taam}}]{Takata2012}
{Takata}, J., {Cheng}, K.~S., \& {Taam}, R.~E. 2012, \apj, 745, 100
\bibitem[Takata et al.(2016)]{Takata2016} 
{Takata}, J., {Ng}, C.~W., \& {Cheng}, K.~S.\ 2016, \mnras, 455, 4249
\bibitem[{{Thompson} \& {Duncan}(1996)}]{Thompson96}
{Thompson}, C., \& {Duncan}, R.~C. 1996, \apj, 473, 322

\bibitem[{{Thompson} {et~al.}(2000){Thompson}, {Duncan}, {Woods},
  {Kouveliotou}, {Finger}, \& {van Paradijs}}]{TD2000}
{Thompson}, C., {Duncan}, R.~C., {Woods}, P.~M., {et~al.} 2000, \apj, 543, 340

\bibitem[{{Watters} {et~al.}(2009){Watters}, {Romani}, {Weltevrede}, \&
  {Johnston}}]{Watters2009}
{Watters}, K.~P., {Romani}, R.~W., {Weltevrede}, P., {et~al.} 2009,
  \apj, 695, 1289
\bibitem[Watters \& Romani(2011)]{WR2011} Watters, K.~P. \& Romani, R.~W.\ 2011, \apj, 727, 123
\bibitem[{{Weltevrede} {et~al.}(2011{\natexlab{a}}){Weltevrede}, {Johnston}, \&
  {Espinoza}}]{WJE2011}
{Weltevrede}, P., {Johnston}, S., \& {Espinoza}, C.~M. 2011{\natexlab{a}},
  \mnras, 411, 1917

\bibitem[{{Weltevrede} {et~al.}(2011{\natexlab{b}}){Weltevrede}, {Johnston}, \&
  {Espinoza}}]{Weltevrede2011}
{Weltevrede}, P., {Johnston}, S., \& {Espinoza}, C.~M. 2011{\natexlab{b}}, \mnras, 411, 1917

\bibitem[{Weng} {et~al.}(2015)]{weng15} Weng, S.-S., Zhang, S.-N., Yi, S.-X., {et~al.}  \ 2015, \mnras, 450, 2915 

\bibitem[{{Woods} \& {Thompson}(2006)}]{WT2006}
{Woods}, P.~M., \& {Thompson}, C. 2006, {Soft gamma repeaters and anomalous
  X-ray pulsars: magnetar candidates}, ed. W.~H.~G. {Lewin} \& M.~{van der
  Klis}, 547--586

\bibitem[{{Woods} {et~al.}(2004){Woods}, {Kaspi}, {Thompson}, {Gavriil},
  {Marshall}, {Chakrabarty}, {Flanagan}, {Heyl}, \& {Hernquist}}]{Woods04}
{Woods}, P.~M., {Kaspi}, V.~M., {Thompson}, C., {et~al.} 2004, \apj, 605, 378

\bibitem[{{Younes} {et~al.}(2016){Younes}, {Kouveliotou}, \&
  {Roberts}}]{Younes2016}
{Younes}, G., {Kouveliotou}, C., \& {Roberts}, O. 2016, GRB Coordinates
  Network, Circular Service, No.~19736, \#1 (2016), 19736

\end{thebibliography}
\end{document}